\documentstyle[12pt,epsf]{article}
\begin{document}
\begin{titlepage}
\begin{flushright}
KUNS-1858\\
\end{flushright}

\begin{center}
\vspace*{10mm}

{\LARGE \bf Dilaton and Moduli Fields\\
\vspace{1.5mm}
in $D$-term inflation}
\vspace{12mm}

{\large
Tatsuo~Kobayashi\footnote{E-mail address:
kobayash@gauge.scphys.kyoto-u.ac.jp}
}
\vspace{6mm}

{\it Department of Physics, Kyoto University,
Kyoto 606-8502, Japan}\\[1mm]


\vspace{6mm}

{\large
Osamu~Seto\footnote{E-mail address: osamu@mail.nctu.edu.tw}}
\vspace{6mm}

{\it Institute of Physics, National Chiao Tung University,
Hsinchu, Taiwan 300 R.O.C.}

\vspace*{15mm}

\begin{abstract}
We investigate the possibility of $D$-term inflation within the
framework of
type I string-inspired models. Although $D$-term inflation model has the
excellent property that it is free from the so-called $\eta$- problem, two
serious
problems appear when we embed $D$-term inflation in string theory,
the magnitude of FI term and the rolling motion of the dilation. In the
present
paper, we analyze the potential of $D$-term inflation in type I inspired
models
and study the behavior of dilaton and twisted moduli fields.
Adopting the non-perturbative superpotential induced by gaugino
condensation,
the twisted moduli can be stabilized. If the dilaton is in a certain 
range, it
evolves very slowly and does not run away to infinity.
Thus $D$-term dominated vacuum energy becomes available for driving
inflation.
By studying the density perturbation generated by the inflation model, we
derive the constraints on model parameters and give some implications on
$D$-term inflation in type I inspired models.
\end{abstract}

\end{center}
\end{titlepage}


\section{Introduction}

In the success of slow-roll inflation, the sufficient flat potential for
the inflaton, which is the inflation driving scalar field, is required.
However, it is not so easy to construct a model of inflation where the
corresponding scalar field has a very flat potential in the framework of
supergravity. Usually, the flatness of the potential is expressed as
\begin{eqnarray}
\epsilon &\equiv & \frac{M_p^2}{2}\left(\frac{V'}{V}\right)^2 \ll 1, \\
\eta &\equiv & M_p^2 \frac{V''}{V} \ll 1 ,
\end{eqnarray}
in terms of the slow roll parameters, where the prime denotes the
derivative with respect to the inflaton and $M_p$ is the reduced Planck
mass. In the case that the vacuum energy is dominated by the $F$-term
during the inflation, supergravity effects produce the soft mass of
inflaton field whose magnitude is the same order of $H$ and spoil the
flatness of potential. In other words, such large inflaton mass gives
$\eta\sim 1$ and violates the slow roll condition, which is often referred to
as the $\eta$ problem.

{}From this point of view, the $D$-term inflation is one of attractive
inflation scenarios within the framework of supersymmetric models
\cite{inflation,D-inf},\footnote{See Ref.~\cite{Casas:1988pa} 
for early study on an inflation model relevant to the $D$-term.} 
because the inflaton does not acquire the mass
squared term of the order of $H^2$ and the flatness of the potential is
preserved.
In the $D$-term inflation scenario, Fayet-Iliopoulos (FI) term is
dominant during the inflation and this vacuum energy causes the
accelerating expansion. The FI term comes from the anomalous $U(1)$
symmetry.
In fact, most of 4D string models have anomalous $U(1)$
symmetry for both heterotic models \cite{DSW,KN}
and type I models \cite{typeI}.
These anomalies can be cancelled by the Green-Schwarz
mechanism, where certain fields transform non-trivially
under anomalous $U(1)$ symmetries \cite{Green:sg}.
Such role is played by the dilaton field $S$ in heterotic
models and
the twisted moduli fields $M$ in type I models.
That generates the FI term.
Thus, the $D$-term inflation seems possible to be realized in string models.

However, there are difficulties in the realization of the $D$-term
inflation based on heterotic models. The first problem is the energy
scale of inflation.
The magnitude of the FI term is given as
\begin{equation}
\xi = \frac{g^2}{192\pi^2}{\rm{Tr}}(Q) M_p^2 ,\label{hetero-xi}
\end{equation}
where $g^2$ is the gauge coupling and ${\rm{Tr}}(Q)$ is a model dependent
constant of the order of ${\mathcal O}(10)-{\mathcal O}(10^2)$ \cite{DSW}.
Eq. (\ref{hetero-xi}) reads
$\sqrt{\xi}/M_p = {\mathcal O}(10^{-1}) - {\mathcal O}(10^{-2})$.
On the other hand, the CMB
anisotropy requires $\xi^{1/2} \simeq 10^{15}$ GeV. We find that the
theoretical prediction is too larger to meet the observational estimation.
This inconsistency is still a serious problem, although several studies to
construct a model with effectively suppressed FI terms have been
done in heterotic models \cite{Espinosa:1998ks,Lesgourgues:1998kj}.
The second problem is due to the dilaton dependence of the FI term and the
anomalous $U(1)$ gauge coupling.
The presence of the dilaton $S$ and several types of moduli
fields is one of important features in superstring theory.
Vacuum expectation values (VEVs) of these fields
determine the magnitudes of all the couplings,
e.g. gauge couplings and Yukawa couplings.
The dilaton field prevents the realization of $D$-term inflation.
Since $g^2 \propto 1/\mathrm{Re}S$, the $D$-term scalar potential, $V_D =
g^2\xi^2/2$, is in proportion to $(\rm{Re}S)^{-3}$.
Hence the dilaton rolls down the potential to the infinity, $\rm{Re}S
\longrightarrow \infty$, and the $D$-term potential energy goes to zero,
$V_D \longrightarrow 0$. This phenomenon is the same as the dilaton runaway
problem in generic string models and
such runaway behavior of the dilaton and moduli fields prevents a viable
inflation \cite{Brustein:nk}.
Even if we adopt type I models rather than heterotic models,
the FI term is dependent on the field which plays a role in
the Green-Schwarz anomaly cancellation mechanism for the anomalous $U(1)$.
Hence, if these fields run away, the $D$-term inflation can not occur
in 4D string models.

In this paper we study behavior of dilaton and moduli fields in
the $D$-term inflation scenario with the above two problems in mind.
The purpose of the present paper is to explore a way to avoid
these problems at the same time.
Actually, studies on the runaway problem have been done
for heterotic models
in Refs. \cite{Matsuda:1997ys,King:1998uv}.
Now, in particular, we will consider type I string-inspired
models. The $D$-term inflation in type I inspired model has been studied in
Ref.\cite{Halyo:1999bq} and it was shown that the magnitude of
FI-term is reducible to a desired value.
This result arise from the facts that the FI-term is determined by the
expectation value of the twisted moduli and the string scale $M_s$ is
independent of the Planck scale $M_p$ \cite{Witten:1996mz}
in type I string models.
In fact, the twisted moduli field plays a role in the Green-Schwarz anomaly
cancellation in 4D type I string models.
However, in Ref.\cite{Halyo:1999bq} 
the stabilization of the twisted moduli and the runway problem 
have not been discussed.
These issues might be not trivial. Indeed, stabilization of the
twisted moduli
fields was studied and different aspects from
those of the dilaton field was revealed \cite{Higaki:2003zk}
\footnote{See also Refs.\cite{Abel:2000tf,Ciesielski:2002fs}.}.
As we will show, the twisted moduli field cannot stabilize with
non-vanishing vacuum energy in only $D$-term potential and any
vacuum energy driving inflation does not appear,
unless we assume a specific K\"{a}hler potential.
Therefore, stabilization of the twisted moduli is also an important issue
for the inflation in Type I models.


This paper is organized as follows.
In the next section, we review briefly the $D$-term inflation.
In section 3, we study behavior of dilaton and twisted moduli fields.
It will be shown that the twisted moduli field
is stabilized during the inflation.
In section 4, dynamics of the inflaton and dilaton fields
are studied, and we will derive constraints on parameters in
our models.
Section 5 is devoted to conclusions and discussions.


\section{$D$-term inflation}

Here we give a brief review on the $D$-term
inflation \cite{inflation,D-inf}.
We consider the N=1 supersymmetric model with $U(1)$ gauge
symmetry and the non-vanishing FI term $\xi$.
The model includes three matter fields $X$ and $\phi_\pm$.
The fields $\phi_\pm$ have $U(1)$ charges $\pm 1$ while
$X$ has no $U(1)$ charge.
The $U(1)$ $D$-term is written as
\begin{equation}
D= \xi +|\phi_+|^2 - |\phi_-|^2 .
\end{equation}
Hereafter we take the charge assignment such that $\xi > 0$.
Suppose that the following superpotential,
\begin{equation}
W = \lambda X \phi_+ \phi_- .
\end{equation}
Then, the scalar potential is written as
\begin{eqnarray}
V &=& \sum_i |\partial_i W|^2 + \frac{g^2}{2}D^2, \\
&=& \lambda^2 |X|^2(|\phi_-|^2 + |\phi_+|^2)
+ \lambda^2|\phi_- \phi_+|^2
+ \frac{g^2}{2}(\xi + |\phi_+|^2 - |\phi_-|^2)^2 .
\end{eqnarray}
The true vacuum of this potential corresponds to
\begin{equation}
X= \phi_+ =0, \qquad |\phi_-| = \sqrt{\xi} ,
\end{equation}
and supersymmetry (SUSY) is not broken.

For a value of $X$ fixed, we analyze the minimum of $V$.
We define
\begin{equation}
X_c \equiv {g \over \lambda} \sqrt{\xi} .
\end{equation}
For $|X| < X_c$, the minimum corresponds to
\begin{equation}
|\phi_-|^2 = \xi- {\lambda^2 X^2 \over g^2}, \qquad
|\phi_+ | = 0 .
\end{equation}
On the other hand, for $|X| > X_c$, the minimum
corresponds to
\begin{equation}
|\phi_\pm| =0 .
\end{equation}
In this case, the vacuum energy is obtained as
\begin{equation}
V = \frac{g^2}{2} \xi^2 ,
\end{equation}
and the radial part of $X$ is identified with the inflaton.
Although the mass of $X$ vanishes at the tree-level, since the
supersymmetry
breaking by the non-vanishing $D$-term generates the mass difference
between the masses of $\phi_\pm$,
\begin{equation}
m^2_\pm = \lambda^2 |X|^2 \pm g^2 \xi
\end{equation}
and the masses of those fermionic partners,
then the one-loop effective potential is given as
\begin{equation}
V_{\rm 1-loop} = {g^2 \over 2} \xi^2 \left(1
+ \frac{g^2}{16 \pi^2}
\ln {\lambda^2 |X|^2 \over \Lambda^2 } \right) ,
\end{equation}
where $\Lambda$ is the renormalization scale.
That generates the mass term of $X$ and the potential is slightly
lifted.
Then, the inflaton slowly rolls down the potential.
{}From the following equation,
\begin{equation}
\eta = -\frac{g^2}{8\pi^2}\frac{M_p^2}{|X|^2} ,
\end{equation}
we find that the slow roll condition is violated $(\eta \sim 1)$ when the
inflaton reach at
\begin{equation}
|X_f|^2 = \frac{g^2}{8\pi^2} M_p^2,
\end{equation}
and the inflation ends.

We have reviewed on the simplest $D$-term inflation model.
However, as noted in Introduction, the magnitude of FI term $\xi$ in such
simplest model is required to be much smaller than the values of $\xi$ derived
from 4D heterotic models.
Ref. \cite{Espinosa:1998ks} has proposed the mechanism
in heterotic models to reduce effectively the FI term
compared with its original value.
Here we also give a brief review on such models.
Instead of the single field $X$, we consider two fields $X_\pm$.
Generic string models have several $U(1)_a$ symmetries.
Some of them are anomalous and others are anomaly-free.
In Ref. \cite{Espinosa:1998ks} 4D heterotic models
were studied and the FI term was considered for only one of
$U(1)$ symmetries.
That is because only one $U(1)$ symmetry can be
anomalous in a 4D heterotic model.
However, we will study inflation models inspired by
type I string models, where
two or more $U(1)$ symmetries can be anomalous.
Thus, here we extend the model of Ref. \cite{Espinosa:1998ks}
into the case with many FI terms, $\xi_a$,
while $\xi_a =0$ for anomaly-free $U(1)_a$.
Suppose that $X_\pm$ have $U(1)_a$ charges $\pm Q_a$.
Then the $D$-term is written as
\begin{equation}
D_a = \xi_a + Q_a(|X_+|^2 - |X_-|^2) + \cdots ,
\end{equation}
where the ellipsis denotes contributions due to other chiral
matter fields.
We consider large field values for $X_\pm$,
$X_\pm \gg \sqrt{\xi_a}$
as before and assume that the other fields gain mass terms,
e.g.
\begin{equation}
W = \lambda \frac{X_+ X_-}{M}\phi_+ \phi_-.
\end{equation}
Then, the matter fields other than $X_\pm$ vanish
for $X_\pm$, $X_\pm \gg \sqrt{\xi_a}$, and the $D$-term is
dominated by $X_\pm$.
In this case, the scalar potential is written as
\begin{equation}
V_D = \sum_a \frac{g_a^2}{2}(\xi_a + Q_a(|X_+|^2 - |X_-|^2) )^2 .
\end{equation}
The direction $|X_+| = |X_-|$ corresponds to
the inflaton.
The minimum of $V_D$ is obtained as
\begin{equation}
(V_D)_{min} \equiv \frac{g^2_{eff}}{2}\xi^2_{eff}
= \sum_a \frac{g^2_a}{2}\xi^2_a -
\frac{(\sum_a g^2 Q_a \xi_a)^2}{\sum_ag^2_aQ_a^2} ,
\end{equation}
for
\begin{equation}
|X_+|^2 - |X_-|^2 = - \frac{\sum_a g^2_a Q_a \xi_a}{\sum_a g^2_a
Q_a^2} .
\end{equation}
In a certain situation (maybe with fine-tuning),
this vacuum energy $(V_D)_{min}$ becomes
much smaller than $O(g^2_a \xi^2_a)$.
Furthermore, in Ref. \cite{Lesgourgues:1998kj} more
fields $X_i$ relevant to the inflaton have been
introduced and in such case
an effective value of FI term can be reduced.


\section{Behavior of dilaton and twisted moduli fields}

In this section we study behavior of
dilaton and twisted moduli fields in
type I string-inspired $D$-term inflation model.
We consider the case that the anomalous $U(1)$
is originated from the $D9$ brane.
In this case the gauge kinetic function is obtained
as \cite{Ibanez:1998rf}
\begin{equation}
f_9 = S + \sigma \frac{M}{M_s},
\end{equation}
where $\sigma$ is a model-dependent constant and 
we define $\tilde \sigma = \sigma /M_s$.
The corresponding gauge coupling $g_9$ is obtained
as $g^2_9 = 1/Re(f_9)$.
The K\"ahler potential of the dilaton filed $S$ is written as
\begin{equation}
K(S,\bar S) = - \ln (S + \bar S).
\end{equation}
On the other hand, the K\"ahler potential of the twisted moduli field
$K(M,\bar M)$ is not clear.
For small values of $M$, it can be expanded as \cite{Poppitz:1998dj}
\begin{equation}
K(M,\bar M) = \frac{1}{2}(M + \bar M)^2 + \cdots .
\end{equation}
However, the reliability of this form is not clear
for large values of $M$, i.e. $M \geq O(1)$.

The twisted moduli field plays a role in the
4D Green-Schwarz anomaly cancellation mechanism, and
the FI term is obtained as
\begin{equation}
\xi = \delta_{GS} K'(M,\bar M) \label{FI},
\end{equation}
where $\delta_{GS}$ is a model-dependent constant.
Thus the $D$-term scalar potential is obtained
during the inflation,
\begin{equation}
V_D = \frac{[\delta_{GS} K'(M,\bar M)]^2}
{S + \bar S + \tilde \sigma (M + \bar M)} .
\label{VD-D9}
\end{equation}

First, let us study behavior of $S$ during the
inflation.
If $S$ run away into infinity, the $D$-term inflation can not succeed.
Here, we introduce the canonically normalized
dilaton field $\phi$, which is defined as
\begin{equation}
\phi = M_p\frac{1}{\sqrt{2}}\ln \frac{S + \bar S}{2} .
\end{equation}
The first and second derivatives of $V_D$ by
$\phi$ are obtained as
\begin{eqnarray}
\frac{\partial V_D}{\partial \phi} &=& -
\frac{\sqrt 2 (S+\bar S)}{S + \bar S + \tilde \sigma
(M + \bar M)}\frac{V_D}{M_p} , \\
\frac{\partial^2 V_D}{\partial \phi^2} &=& -
\frac{2\sqrt 2 (S+\bar S)}{S + \bar S + \tilde \sigma
(M + \bar M)}\frac{V_D}{M_p^2} \left(1 - \frac{2 (S+\bar S)}
{S + \bar S + \tilde \sigma (M + \bar M)}\right) .
\end{eqnarray}
Suppose the twisted moduli $M$ is stabilized somehow
such that $S + \bar S \ll \tilde \sigma (M + \bar M)$.
Then, we obtain
\begin{eqnarray}
\left( \frac{1}{V_D}\frac{\partial V_D}{\partial \phi}
\right)^2 &\simeq & 2 \left(\frac{S+\bar S}{\tilde \sigma(M + \bar M)}
\right) \ll 1, \\
\frac{1}{V_D}\frac{\partial^2 V_D}{\partial \phi^2}
&\simeq & - \frac{2 \sqrt 2(S + \bar S)}{\tilde \sigma (M + \bar M)}
\ll 1,
\end{eqnarray}
that is, the slow roll condition is satisfied for the
dilaton field and it does not run away.
Note that this result is independent of the form of
$K(M,\bar M)$.

We have considered the case that the $U(1)$ gauge
multiplet is originated from the $D9$ brane.
We can easily extend the above analysis into another case.
For example, when the $U(1)$ gauge multiplet is originated
from the $D5$ brane wrapping a 2D torus of
the 6D compact space.
In this case, the gauge kinetic function is obtained as
\begin{equation}
f_{5i} = T_i + \tilde \sigma (M + \bar M) ,
\end{equation}
where $T_i$ is the moduli field whose VEV determines
the volume of the 2D torus.
Its K\"ahler potential is the same as the dilaton field,
i.e.,
\begin{equation}
K(T_i,\bar T_i) = - \ln (T_i + \bar T_i) .
\end{equation}
Hence, the $D$-term scalar potential during
the $D$-term inflation is written as
\begin{equation}
V_D = \frac{[\delta_{GS} K'(M,\bar M)]^2}
{T_i + \bar T_i + \tilde \sigma (M + \bar M)} .
\end{equation}
In a similar way to the dilaton field,
the moduli field $T_i$ does not run away if
$M$ is stabilized such that $T_i + \bar T_i \ll
\tilde \sigma (M + \bar M)$.


Next, we study the stabilization of $M$ in the model
with $V_D$, Eq. (\ref{VD-D9}),
that $U(1)$ is originated from the $D9$ brane.
For $S + \bar S \ll \tilde \sigma (M + \bar M)$, the potential
reduces to
\begin{equation}
V_D= \frac{[\delta_{GS} K'(M,\bar M)]^2}
{\tilde \sigma (M + \bar M)} .
\end{equation}
Unfortunately, only the term $V_D$ does not stabilize
the twisted moduli $M$ at a finite value with non-vanishing vacuum energy,
unless we choose a special form of $K(M,\bar M)$.
In fact, the minimum in this potential is located at $M+\bar M=0$ and
$\left.V_D\right|_{M+\bar M=0}=0$ for the K\"{a}hler
potential, $K(M,\bar M) = (M + \bar M)^2/2$.
Hence, another term is necessary for stabilization of $M$
such that the $D$-term inflation can be realized.
We assume that gaugino condensation of another
non-abelian gauge group generates the
following non-perturbative superpotential,
\begin{equation}
W_{np} = d e^{-\Delta (S + \tilde \sigma' M)},
\end{equation}
where $\Delta = 24 \pi^2/(-b)$ and $b$ is the one-loop
beta-function coefficient of the gauge coupling.
In general, the constant $\sigma'$ is different from $\sigma$.
Anyway one needs a stabilization mechanism of $S$ as well as $M$
in the true vacuum after the inflation ends.
It is the usual approach to stabilize $S$ by this type of
non-perturbative superpotential.
Thus, it is natural to assume the above superpotential.
For stabilization of $S$ in the true vacuum, double or more gaugino
condensations
are often used in the so-called race-track model.
Here, for simplicity, we consider the case that only the single gaugino
condensation potential is dominant during the inflation.
As we did in estimation of $V_D$, we assume
$S+\bar S \ll \tilde \sigma (M + \bar M)$ during inflation, that is,
we have $W_{np} = d e^{-\Delta \tilde \sigma' M}$.
In this case, the scalar potential in the global SUSY is
written as \footnote{When we study the scalar potential
within the
framework of supergravity theory, we obtain the qualitatively
same result.}
\begin{equation}
V = (\Delta \tilde \sigma')^2 d^2 e^{-\Delta \tilde \sigma' m} +
\frac{\delta_{GS}^2}{\tilde \sigma }m,
\label{V-inf}
\end{equation}
where $m \equiv M + \bar M$.
We need the explicit form of $K(M,\bar M)$ to discuss
the stabilization of $M$. Here and hereafter we take
\begin{equation}
K(M,\bar M) = \frac{1}{2}(M + \bar M)^2. \label{M,Kahler}
\end{equation}
The stationary condition $\frac{\partial V}{\partial m} =0$
is satisfied by
\begin{equation}
\langle M+\bar M\rangle_{inf} =
\frac{1}{\Delta \tilde \sigma'}\ln
\left(\frac{\tilde \sigma d^2(\Delta \tilde \sigma')^3}{\delta_{GS}^2}
\right) ,
\end{equation}
where $\langle ... \rangle_{inf}$ denotes the expectation value during
inflation.
If the effective mass squared during inflation,
\begin{equation}
\left.\frac{\partial^2 V}{\partial m^2}\right|_{m=\langle M+\bar
M\rangle_{inf}} \simeq \frac{\Delta \tilde \sigma}{\langle M+\bar
M\rangle_{inf}}V_D = 3\frac{\Delta \tilde \sigma}{\langle M+\bar
M\rangle_{inf}}{M_p}^2H^2 ,
\end{equation}
is enough larger than $H^2$ around this stationary point,
the twisted moduli can be stabilized. The condition is
satisfied in the present model because of $\Delta \tilde \sigma M^2_s/
\langle M+\bar M\rangle_{inf}\gg 1$ and $M_p\sim M_s$
\footnote{This relation is derived in the next section.}, as we will see
soon.
Note that the key-point for stabilization is the polynomial form
of the K\"ahler potential, in particular the canonical form.
When we consider additional higher terms $(M + \bar M)^n$ ($n > 2$)
in the K\"ahler potential, we obtain qualitatively similar
result for stabilization.
However, if the twisted moduli field has the logarithmic form
of K\"ahler potential, this type of stabilization can not
be realized.

Now let us estimate $\langle M+\bar M\rangle_{inf}$.
If $b/ \sigma' ={\mathcal O}(1)$, we estimate
$(\Delta \sigma')^{-1} = {\mathcal O}(10^{-3})-{\mathcal O}(10^{-2})$,
because of $(\Delta \sigma')^{-1} = 0.0004 \times b /\sigma'$.
For such small value of
$(\Delta \sigma')^{-1} = {\mathcal O}(10^{-3})- {\mathcal O}(10^{-2})$,
the factor $(\Delta\sigma')^3$ inside of the logarithmic
function enlarges $\langle M+\bar M\rangle_{inf}$
by a factor of $O(10)$.
Also, the GS coefficient $\delta_{GS}$ may enlarge
$\langle M+\bar M\rangle_{inf}$
by a few factor for $\delta_{GS} \ll 1$.
Then we can estimate
\begin{equation}
\frac{\langle M+\bar M\rangle_{inf}}{M_s} 
= {\mathcal O}(10^{-2})-{\mathcal O}(10^{-1}) .
\label{value-m}
\end{equation}
Fig.~1 shows $V$ against $m$ for $\Delta \sigma' = 100$ and
$\delta_{GS}^2/\sigma d^2 = 0.01 $.
The form of K\"ahler potential, Eq.(\ref{M,Kahler})
is not reliable for a large value of
$\langle M+\bar M\rangle_{inf}$, i.e.
$\langle M+\bar M\rangle_{inf}/M_s
\geq {\mathcal O}(1)$. From this point, the above result that $M$ is
stabilized as a small value is favorable.
Another nice point of the result is that the stabilized value
satisfies
\begin{equation}
\langle M+\bar M\rangle_{inf} \gg \frac{1}{\Delta \tilde 
\sigma'} , \label{M_inf}
\end{equation}
when the value inside of the logarithmic function is large.
Note that $F$-term and $D$-term parts in the scalar potential
satisfy
\begin{equation}
\frac{V_D}{V_F} \simeq \Delta \tilde \sigma'\langle M+
\bar M\rangle_{inf} . \label{DtoF}
\end{equation}
Eqs.(\ref{M_inf}) and (\ref{DtoF}) imply the following inequality,
\begin{equation}
V_F \ll V_D ,
\end{equation}
that is, with this stabilized value of
$\langle M+\bar M\rangle_{inf}$
the $D$-term scalar potential is dominant.
That allows to realize the $D$-term inflation.
Therefore, the vacuum energy during inflation is estimated as
\begin{equation}
V = \frac{\delta_{GS}^2}{\tilde \sigma} \langle M+\bar M\rangle_{inf} .
\end{equation}

In explicit type I string models through type IIB orientifold
construction,
the GS coefficient $\delta_{GS}$ is calculated
in the unit of $M_s$ as
$\delta_{GS}/M_s =
O(10^{-1}) - O(10^{-3})$ \cite{Antoniadis:2002cs}.
It is not completely clear what is a natural value of
$\delta_{GS}$ from the viewpoint of generic type I models
except type IIB construction.
Furthermore, as reviewed in the previous section,
the original value of the FI term appears as the
inflation vacuum energy in the simplest model,
but it can be effectively reduced in the models with
more fields and $U(1)$ symmetries.
As another remark, if $d$ is suppressed by any mechanism, the stabilized
value $\langle M+\bar M\rangle_{inf}/M_s$ becomes small
like $O(10^{-2}) - O(10^{-3})$.

\begin{figure}
\epsfxsize=0.7\textwidth
\centerline{\epsfbox{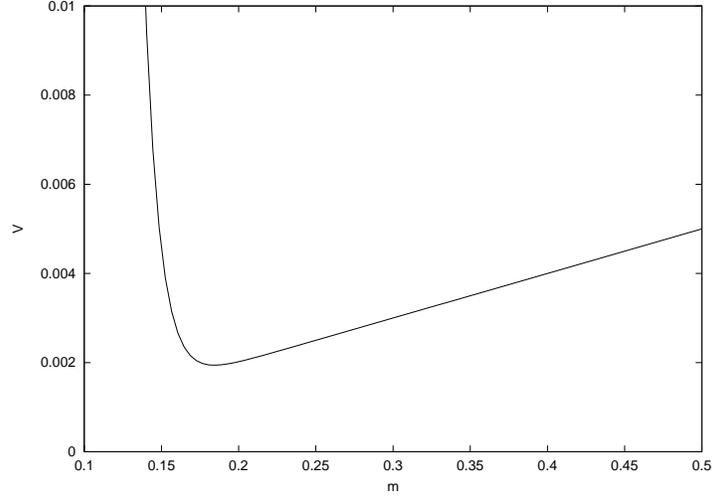}}
\caption{The scalar potential (\ref{V-inf}) of $m$ for $\Delta \sigma' =
100$
and $\delta_{GS}^2/\sigma d^2 = 0.01 $.}
\label{fig1}
\end{figure}
%


\section{Dynamics and density perturbation}

Now let us discuss the dynamics in this inflation model.
The relevant potential including one-loop correction during inflaton is
given by
\begin{eqnarray}
V &=& \frac{\delta_{GS}^2\langle M+\bar M\rangle_{inf}^2}{S+\bar
S+\tilde \sigma\langle M+\bar
M\rangle_{inf}} \nonumber \\
&& \times \left[1+\frac{1}{16\pi^2}\frac{2}{S+\bar S+\tilde \sigma\langle
M+\bar M\rangle_{inf}}\ln\frac{\lambda^2|X|^2}{(S+\bar S)\Lambda^2}\right] .
\label{potential}
\end{eqnarray}
Here the $F$-term contribution is negligibly smaller than the $D$-term in
potential, as already mentioned. On the other hand, one should notice
that the stabilization of twisted moduli is achieved by the $F$-term
potential and we can replace $M+\bar M$ with $\langle M+\bar
M\rangle_{inf}$. Furthermore, hereafter we will consider only the
region, $S+\bar S \ll \tilde \sigma\langle M+\bar M\rangle_{inf}$ in order to
avoid the runway motion of the dilaton and obtain a successful inflation.
In this sence, the initial value problem of the dilaton,
why the dilaton takes such a initial value, remains unsolved.

{}From
\begin{equation}
\eta_{XX} = -\frac{1}{8\pi^2}\frac{2}{\tilde \sigma\langle M+\bar
M\rangle_{inf}}\frac{M_p^2}{X^2},
\end{equation}
we find that when the inflaton, $X$,\footnote{Hereafter, we omit the
absolute value symbol, $| |$.} reaches at the critical point, $X_f$,
where
\begin{equation}
X_f^2 = \frac{1}{8\pi^2}\frac{2M_p^2}{\tilde \sigma\langle M
+\bar M\rangle_{inf}},
\end{equation}
the slow roll condition is violated. On the other hand, $X_c$ is
expressed as
\begin{equation}
X_c^2 \simeq \frac{2}{\tilde \sigma\langle M+\bar
M\rangle_{inf}}\frac{\delta_{GS}\langle M+\bar M\rangle_{inf}}{\lambda^2}.
\end{equation}
Hence, if the inequality
\begin{equation}
\frac{M_p^2}{8\pi^2} > \frac{\delta_{GS}\langle M+\bar
M\rangle_{inf}}{\lambda^2}
\end{equation}
hold, inflationary expansion terminates when inflaton reaches at $X_f$
rather than $X_c$. This inequality holds for $\lambda \sim 1$.

Then the equations of motion for the homogeneous part of scalar fields,
$X$ and $\phi$, are respectively given by
\begin{eqnarray}
\ddot{X}+ 3H\dot{X} &=& -\partial_{X}V \nonumber \\
& \simeq& -\frac{\delta_{GS}^2}{\tilde \sigma^2}\frac{1}{8\pi^2}\frac{2}{X} ,
\end{eqnarray}
and
\begin{eqnarray}
\ddot{\phi}+ 3H\dot{\phi} &=& -\partial_{\phi}V \nonumber \\
& \simeq &
\frac{\delta_{GS}^2}{\tilde \sigma^2}\frac{2^{3/2}}{M_p}e^{\sqrt{2}\phi/M_p} ,
\end{eqnarray}
where we ignored the dilaton dependence in the radiative correction part
because of the weakness of the dependence. The solutions under the slow
roll approximation are
\begin{eqnarray}
X^2 &=& X_e^2
+\frac{\delta_{GS}^2}{\tilde \sigma^2}\frac{1}{2\pi^2}\frac{H(t_e-t)}{3H^2} ,
\label{solution-X} \\
e^{-\sqrt{2}\phi/M_p} &=& e^{-\sqrt{2}\phi_e/M_p} +
\frac{4\delta_{GS}^2}{\tilde \sigma^2}\frac{H(t_e-t)}{3M_p^2H^2}
\label{solution-phi} ,
\end{eqnarray}
with $X_e = X(t_e)$ and $\phi_e = \phi(t_e)$, where $t_e$ denotes the
time of the end of the inflation. By using the definition of the number
of $e$-folds $-d N \equiv H dt $, Eq.(\ref{solution-phi}) is rewritten as
\begin{equation}
\left.\frac{S+\bar S}{\tilde \sigma\langle M+\bar M\rangle_{inf}}\right|_{t_e}
= \frac{1}{\frac{\tilde \sigma\langle M+\bar M\rangle_{inf}}{S+\bar S} -2 N}.
\end{equation}
Since the l.h.s. must be less than the order of $10^{-1}$, we find that
$\tilde \sigma\langle M+\bar M\rangle_{inf} /(S+\bar S) \sim 10^2$ 
for the present
horizon scale.
Similarly, from Eq.(\ref{solution-X}) we obtain
\begin{equation}
X^2 = X_e^2 + \frac{N}{2\pi^2\tilde \sigma \langle M+\bar
  M\rangle_{inf}}
M_p^2 ,
\label{Xsol}
\end{equation}
we unfortunately found the somewhat large field value region,
$X = {\mathcal O}(1)M_p$, is used in this inflation model.
This is the undesirable feature in this model.
Here we estimate the following quantity
\begin{eqnarray}
\left(\frac{\dot{\phi}}{\dot{X}}\right)^2 &=& (8\pi^2\tilde\sigma\langle
M+\bar M\rangle_{inf})^2\left(\frac{S+\bar S}{\tilde\sigma\langle M+\bar
M\rangle_{inf}}\right)^2\frac{X^2}{2M_p^2} \nonumber \\
&\simeq& 10^{-1}\left(\frac{\tilde\sigma\langle M+\bar
M\rangle_{inf}}{10^{-1}}\right)
\left(\frac{\left.\frac{S+\bar S}{\tilde\sigma\langle M+\bar
M\rangle_{inf}}\right|_{t_e}}{10^{-1}}\right)^2 \nonumber \\
&& \times \frac{1+2H(t_e-t)}{\left(1+\left.\frac{S+\bar S}{\tilde\sigma\langle
M+\bar M\rangle_{inf}}\right|_{t_e}2H(t_e-t)\right)^2} ,
\end{eqnarray}
where we adopted $X_f$ as $X_e$ and suppose $\sigma \sim 1$.
Thus we find that the ratio of kinetic energies of these scalar fields
takes
a value of ${\mathcal O}(10^{-1})-{\mathcal O}(10^{-2})$, because the
final part
in this expression is almost unity during the inflation.

Next, we turn to the density perturbation generated by this inflation
model. The WMAP data shows that the adiabatic fluctuation is favored
\cite{Bennett:2003bz,Peiris:2003ff}. On the other hand, in general,
a inflation model which contains several evolving scalar fields
produces also the isocurvature fluctuation. Accordingly, we estimate
the contribution from the isocurvature mode and confirm this model
to be consistent with the observations.
The perturbed Einstein equation in Fourier space reads
\begin{eqnarray}
&& \ddot{\delta\phi}+3H\dot{\delta\phi}
+\left(\frac{k^2}{a^2}+V,_{\phi\phi}\right)\delta\phi+V,_{\phi X}\delta
X = -4\dot{\phi}\dot{\Phi} + 2V,_{\phi}\Phi , \\
&& \ddot{\delta X}+3H\dot{\delta X}
+\left(\frac{k^2}{a^2}+V,_{XX}\right)\delta X+V,_{X\phi}\delta\phi =
-4\dot{X}\dot{\Phi} + 2V,_X\Phi , \\
&& \dot{\Phi}+H\Phi =
-\frac{1}{2M_p^2}(\dot{\phi}\delta\phi+\dot{X}\delta X) ,
\end{eqnarray}
where $k$ is the wave number, $a$ is the scale factor, the comma denotes
the derivative of $V$ with respect to the fields $(X,\phi)$,
$\delta\phi$ and $\delta X$ represent the perturbations of scalar
fields $\phi$ and $X$, respectively, and $\Phi$ is the curvature
perturbation variable in the notation of Ref. \cite{Kodama:bj}.
These equations have analytic solutions for a long wavelength under the
slow-roll approximation \cite{Mukhanov:1997fw} ;
\begin{eqnarray}
\delta X &=& (\ln V),_X \left[Q_1 + Q_3 \int_{t_k}^t (\ln(\ln
V),_X),_{\phi} J d\phi\right] , \label{X-sol}\\
\delta\phi &=& (\ln V),_{\phi}\left[Q_1+Q_3 - Q_3 \int_{t_k}^t (\ln(\ln
V),_{\phi}),_X J dX\right] ,\label{phi-sol}
\end{eqnarray}
with
\begin{equation}
J = \exp\left( -\int_{t_k}^t \left[(\ln(\ln V),_X),_{\phi}d\phi +
(\ln(\ln V),_{\phi}),_XdX\right] \right) .
\end{equation}
Here $t_k$ denotes the horizon crossing time for a wave number $k$,
$k=a(t_k)H$. 
In addition, $Q_1$ and $Q_2$ are integration constants and
expressed as
\begin{eqnarray}
Q_1 &=& -\left.\frac{H^2}{\sqrt{2k^3}\dot{X}} e_{X}(\bf{k})\right|_{t_k}
, \\
Q_3 + Q_1&=& -\left.\frac{H^2}{\sqrt{2k^3}\dot{\phi}}
e_{\phi}(\bf{k})\right|_{t_k} ,
\end{eqnarray}
where $e_{X}(\bf{k})$ and $e_{\phi}(\bf{k})$ are classical random
quantities which are normalized as
\begin{equation}
\langle e_X({\mathbf k}) e^*_X({\mathbf k}')\rangle = \langle
e_{\phi}({\mathbf k}) e^*_{\phi}({\mathbf k}') \rangle =
\delta^{(3)}({\mathbf k}-{\mathbf k}'). \nonumber
\end{equation}
The appearance of $Q_3$ is due to the existence of isocurvature perturbation.
Furthermore, as we already mentioned, since the dilaton dependence in the
radiative correction part is negligible, the potential,
Eq.(\ref{potential}), seems a separated form $V = V_{\phi}(\phi)V_X(X)$.
Since $J=1$ in this case, Eqs.(\ref{X-sol}) and (\ref{phi-sol}) can be
rewritten
as
\begin{eqnarray}
\delta X &=& \frac{V,_X}{V}Q_1 , \\
\delta\phi &=& \frac{V,_{\phi}}{V}(Q_1+Q_3) .
\end{eqnarray}
Then, the curvature perturbation is expressed as
\begin{eqnarray}
\Phi &=& -\frac{\dot{H}}{H^2}Q_1 + \frac{(V,_{\phi})^2}{2V^2}Q_3 \\
&=& \left.\frac{\dot{H}}{H^2}\frac{H^2}{\sqrt{2k^3}\dot{X}}
e_{X}({\mathbf k})\right|_{t_k} +
\left.\frac{(V,_{\phi})^2}{2V^2}\left(\frac{H^2}{\sqrt{2k^3}\dot{X}}
e_{X}({\mathbf k})-\frac{H^2}{\sqrt{2k^3}\dot{\phi}}e_{\phi}
({\mathbf k})\right)\right|_{t_k} .
\end{eqnarray}
{}From $ \dot{X}^2 \gg \dot{\phi}^2$, we find the first term is much
larger than the second term and the primary contribution to the density
perturbation comes from the adiabatic fluctuation generated by the
inflaton. Hence, we find that the density perturbation in this
inflation model is almost adiabatic with small isocurvature fluctuations.

Here we shall introduce the so-called Bardeen parameter which is defined by
\begin{equation}
\zeta \equiv \Phi+\frac{2}{3}\frac{\Phi+H^{-1}\dot{\Phi}}{1+w} ,
\end{equation}
which is conserved in the superhorizon scale, $k \ll aH$, if the
fluctuation is adiabatic \cite{Bardeen:qw}.
Here $w$ denotes the ratio of pressure to
energy density. However, the present model contains two slow-rolling
scalar fields and may produce the isocurvature fluctuation. In this
case, the time variation of $\zeta$ is caused by the isocurvature
fluctuation and the Bardeen parameter is no logner conserved quantity on
superhorizon scales.
Indeed, the time variation of $\zeta$ is expressed as
\begin{equation}
\frac{3}{2}H(1+w)\dot{\zeta} =
-c_s^2\frac{k^2}{a^2}\Phi-\frac{\rho}{2}w\Gamma ,
\label{zeta-dot}
\end{equation}
where $c_s^2 = \dot{p}/\dot{\rho}$ and the following quantity 
\begin{equation}
\Gamma = \frac{\delta p}{p} -\frac{c_s^2}{w}\frac{\delta \rho}{\rho}
\end{equation}
represents the amplitude of an entropy perturbation.
In the case that the universe is filled with two component scalar fields
$(X,\phi)$, Eq.(\ref{zeta-dot}) is rewritten as \cite{Garcia-Bellido:1995fz}
\begin{eqnarray}
\dot{\zeta} &=&
\frac{H}{\dot{H}}\frac{k^2}{a^2}\Phi+\frac{H}{2}\frac{\partial}{\partial
t}\left(\frac{\dot{X}^2-\dot{\phi}^2}{\dot{X}^2+\dot{\phi}^2}\right)
\left(\frac{\delta X}{\dot{X}}-\frac{\delta\phi}{\dot{\phi}}\right).
\label{zeta-dot-2}
\end{eqnarray}
In a single field inflation model, either $\dot{X}$ or $\dot{\phi}$ goes
to zero
and the second term in r.h.s of Eq.(\ref{zeta-dot-2}) vanishes. Thus,
$\zeta$
becomes the conserved quantity on superhorizon scales in that case.
However, we need take the variation of $\zeta$ into account,
because there are two evolving scalar fields in the present model.
The spectrum of the density perturbations, $\zeta$, at the horizon
crossing time is
\begin{equation}
\left.\mathcal{P}_{\zeta}\right|_{t_k} =
\left.\left(\frac{H^2}{2\pi|\dot{X}|}\right)^2
\left[1-\frac{1}{2}\frac{\epsilon_{\phi}}{\epsilon_{X}}\right]\right|_{t_k}
,
\end{equation}
up to the first order of $\dot{\phi}^2/\dot{X}^2$, where
$\epsilon_{\phi}= \dot{\phi}^2/(2V)$ and $\epsilon_{X}= \dot{X}^2/(2V)$
are slow roll parameters.
{}From Eq.(\ref{zeta-dot-2}), the variation of the Bardeen parameter after
the horizon crossing is estimated as
\begin{eqnarray}
\Delta\zeta \equiv \int_{t_k}^{t_e}\dot{\zeta}dt =
Q_3\left(\left.\frac{\dot{\phi}^2}{\dot{X}^2}\right|_{t_k}-
\left.\frac{\dot{\phi}^2}{\dot{X}^2}\right|_{t_e}\right) .
\end{eqnarray}
Furthermore, the power spectrum of the density perturbations at the end
of inflation \cite{Garcia-Bellido:1995qq} is estimated as
\begin{equation}
\left.{\mathcal P}_{\zeta}\right|_{t_e} =
\left.\left(\frac{H^2}{2\pi |\dot{X}|}\right)^2\right|_{t_k}
\left[1-2\left.\frac{\epsilon_{\phi}}{\epsilon_{X}}\right|_{t_e}+
\frac{\left(\left.\epsilon_{\phi}/\epsilon_{X}\right|_{t_e}\right)^2}
{\left.\epsilon_{\phi}/\epsilon_{X}\right|_{t_k}}\right] .
\end{equation}
We find that the variation of $\zeta$ on super-horizon scales by
the isocurvature perturbation is suppressed by the power of
$\dot{\phi}^2/\dot{X}^2 = O(10^{-1})-O(10^{-2})$. The amplitude of
curvature perturbation is rewritten as
\begin{eqnarray}
\left.\frac{H^2}{2\pi |\dot{X}|}\right|_{t_k} =
2\frac{\delta_{GS}\langle M+\bar M\rangle_{inf}}{M_p^2}\sqrt{\frac{N}{6}} ,
\end{eqnarray}
and we obtain the constraint on the model parameters
\begin{equation}
\frac{\delta_{GS}\langle M+\bar M\rangle_{inf}}{M_p^2}\simeq 10^{-5}
\label{scalar-constr}
\end{equation}
for the present horizon scale.
For the GS coefficient $\delta_{GS}$, a suppressed value is required
like the heterotic case.
Explicit models through type IIB orientifold construction
seem to derive $\delta_{GS}/M_s = {\mathcal O}(10^{-1}) -
{\mathcal O}(10^{-3})$ \cite{Antoniadis:2002cs}.
Anyway, the GS coefficient $\delta_{GS}$ is model-dependent.
If we find the model that $\delta_{GS}$ is small enough like
$\delta_{GS}/M_s \sim 10^{-3}$,
the string scale can be
almost comparable with $M_p$ or slightly smaller.
Otherwise, if the original FI terms are not
small enough, we would need the mechanism to reduce the effective FI term
as mentioned in Section 2 or a lower string scale.

On the other hand, the spectrum of gravitational wave perturbation
produced by the inflation is given by ${\mathcal P}_g = |H/(2\pi M_p)|^2$
and its observational results show ${\mathcal P}_g \le 10^{-10}$. This
constraint is expressed as $V \le 10^{-8}M_p^4$ in term of the potential.
Therefore we obtain
\begin{eqnarray}
V = \frac{\delta_{GS}^2\langle M +\bar M\rangle_{inf}}{\tilde \sigma} \le
10^{-8}M_p^4 \label{tensor-constr} .
\end{eqnarray}
By combining Eqs.(\ref{scalar-constr}) and (\ref{tensor-constr}), we
obtain $\tilde \sigma\langle M +\bar M\rangle_{inf} \ge 10^{-2}$.
That can be satisfied by a stabilized value (\ref{value-m}) of
$\langle M +\bar M\rangle_{inf}$, e.g.
$\langle M +\bar M\rangle_{inf} \sim 0.1\times M_s$ for $\sigma \sim 1$.

Finally we discuss the evolution of the universe after inflation. The
relevant potential after inflation is given by
\begin{eqnarray}
V &=& \mathrm{(gaugino\,~ condensation\,~ potential )} \nonumber \\
& & +\lambda^2 |X|^2|\phi_-|^2 + \frac{(\delta_{GS}K'(M,\bar M) -
|\phi_-|^2)^2}{S+\bar S+\tilde \sigma(M+\bar M)} .
\end{eqnarray}
After the inflation, scalar fields, $X$ and $\phi_-$ oscillate around their
corresponding minimum $X=0$ and $\langle\phi_-\rangle = \sqrt{\xi}$,
respectively.
Here we should notice that $\xi$ is function of the twisted moduli $M$ and
still a time variant in the reheating stage,
because the twisted moduli does not yet reach at the true
vacuum at this moment. This is a feature which the simplest model of
$D$-term inflation does not possess.

Now, we consider the reheating process according as
Ref. \cite{Kolda:1998kc}. Since there are fields which are charged under
both anomalous $U(1)$ and each of the subgroups of the standard model,
the $D$-term receives a contribution from these field.
As the result, the $D$-term takes the form of
\begin{eqnarray}
D &=& g\left(|\phi_+|^2-|\phi_-|^2+\sum_i q_i|Q_i|^2 + \xi\right) ,\\
\xi &=& \delta_{GS}K'(M,\bar M) ,
\end{eqnarray}
where $Q_i$ are charged fields for both anomalous $U(1)$ and gauge groups of
the standard model and $q_i$ are their positive charge. In the case that
 fields $\varphi_i$ of the minimal supersymmetry standard model 
belong to the fields $Q_i$, the $\phi_-$ field couples
with them by the interaction,
\begin{equation}
{\mathcal L}{}_{int} = g^2\sqrt{\xi}\sum_i q_i|\varphi_i|^2(\delta\phi_-
+ \delta\phi^{\dagger}_-),
\end{equation}
where $\delta\phi_-\equiv \phi_- -\langle\phi_-\rangle$. Hence the $\phi_-$
field
would immediately decay through this gauge interaction.
On the other hand, the decay of $X$
would delay. The mass of $\phi_-$ is $m_{\phi_-} = (g/\lambda)m_X$.
Since $g > 1$ in our model at the time of the end of inflation, the decay
of $X$ to $\phi_-$ would be forbidden kinematically. As a results, $X$
decays to
light matter through higher-dimensional interactions and the reheating can
be completed with an enough high reheating temperature.

In this scenario, huge entropy after the reheating by the inflaton is
produced by the decay of the dilaton and moduli fields in late time.
Therefore, this $D$-term inflation model should be incorporated with
the Affleck-Dine baryogenesis \cite{Affleck:1984fy}. The Affleck-Dine
baryogenesis with the dilution by the dilaton and moduli decay have been
investigated \cite{Moroi:1994rs,Dolgov:2002vf}. However, the
actual estimation of the baryon to entropy ratio is beyond the scope of
this paper, because their full potential strongly depend on the form of the
K\"{a}hler potential of twisted moduli and the condensation gauge
groups \cite{Higaki:2003zk}.

Furthermore, it is pointed out that cosmic strings generated after
$D$-term inflation can modify the CMB anisotropy significantly
\cite{Endo:2003fr}.
Huge entropy production in a late time is desirable from this point of
view,
because the dilution by the entropy production is one possible way to
remove
the effect by cosmic strings.
Moreover there is one thing to note.
The FI term is given by Eq.(\ref{FI}) and its magnitude
is parameterized by the expectation value of the twisted moduli.
Therefore the tension of cosmic strings in late time is not necessarily
the same
as the scale of FI term during inflation, namely,
$\langle M+\bar M\rangle_{inf} \neq \langle M+\bar M\rangle$,
where $\langle ... \rangle$ denotes the VEV.
This feature might change the situation of the effect by cosmic strings.


\section{Conclusions and Discussions}

We have studied the $D$-term inflation within the framework of
type I string-inspired models.
In this case, the twisted moduli field plays a significant role.
For stabilization of the twisted moduli, the polynomial
form of the K\"ahler potential is important.
The stabilization of the twisted moduli during inflation is achieved with
the help of the gaugino condensation potential and this stabilization
mechanism
does not spoil the situation that the potential is dominated by $D$-term.
Furthermore it is a favorable property that the expectation value of the
twisted
moduli during inflation is enough smaller than the unity where the
K\"{a}hler
potential $K=(M +\bar M)^2/2$ is valid, while this result leads to one
shortcoming that
the inflaton $X$ must take a somewhat large value $X > M_p$ during
inflation.

At first, we found that the magnitude of FI term would be reducible to a
desirable value within this framework even if we do not assume 
a hierarchically large 
gap between the string scale $M_s$ and the Planck scale $M_p$.

Concerning the second problem, or dilaton runaway problem,
we found the condition to avoid this difficulty for the initial value of
the
dilaton.
In the potential with the fixed twisted moduli, we could find that the
field
region where the dilaton does not run away, $S+\bar S\ll \tilde
\sigma\langle M+\bar
M\rangle_{inf}$. If the dilaton takes such initial value at the
pre-inflation
stage, it is possible that the universe undergoes the quasi de-Sitter
expansion,
because the dilaton can evolve slowly.
Of course, at the moment, we cannot answer the question why the
dilaton takes
such initial value.
The point we would like to emphasize in the present paper is that $D$-term
inflaton is possible in type I string-inspired model under certain conditions.
On the other hand, the small expectation value of the dilaton during
inflation
might be a good point from another point of view. In studies on the
evolution
of the dilaton in string cosmology,
the initial condition that the dilaton takes a smaller value than the
VEV at
the beginning, $\left.(S+\bar S)\right|_{t_i} <
\langle S+\bar S\rangle$, has been adopted
\cite{Brustein:nk,Dolgov:2002vf,Barreiro:1998aj}.
Therefore, the initial value of the dilaton set by this inflation is
consistent
with the initial condition in these studies. Although these are based on
a heterotic model, the situation would not change essentially.
That might lead to the dilaton stabilization at the correct vacuum.

Moreover, the ingenious point in this model is that the condition which
prohibits the dilaton from rolling down the potential also suppresses
the growth of the unnecessary isocurvature fluctuation.
Thus, we can obtain the almost adiabatic density perturbation.

What we have to investigate further is the late time evolution of
the universe. In the present paper, we have not estimated the
baryogenesis
and the effect of cosmic strings and so on, because they would involve
us in
other factors such as a concrete potential for the supersymmetry breaking
than the potential of inflation. These are important issues in future works.


\section*{Acknowledgment}

T.~K. is supported in part by the Grant-in-Aid for
Scientific Research from Ministry of Education, Science,
Sports and Culture of Japan (\#14540256).
O.~S. is supported by National Science Council.


\end{document}